\RequirePackage{snapshot}
\pdfoutput=1
\documentclass[a4paper,11pt]{article}

\usepackage{url}

\usepackage{fullpage}

\usepackage{amsmath,amsthm,amssymb}
\usepackage{xcolor}
\usepackage{graphicx}
\usepackage{bbold}
\usepackage{textcomp} 
\usepackage{siunitx}

\usepackage{authblk}
\renewcommand\Affilfont{\itshape\small}

\renewcommand{\vec}[1]{\boldsymbol{#1}}

\newcommand{\vnabla}{\vec{\nabla}}

\title{
  A self-consistent spin-diffusion model for micromagnetics
}

\author[1]{Claas Abert\thanks{claas.abert@tuwien.ac.at}}
\author[2]{Michele Ruggeri}
\author[1]{Florian Bruckner}
\author[3]{Christoph Vogler}
\author[4]{Aurelien Manchon}
\author[2]{Dirk Praetorius}
\author[1]{Dieter Suess}
\affil[1]{Christian Doppler Laboratory of Advanced Magnetic Sensing and Materials, Institute of Solid State Physics, TU Wien, Austria}
\affil[2]{Institute for Analysis and Scientific Computing, TU Wien, Austria}
\affil[3]{Institute of Solid State Physics, TU Wien, Austria}
\affil[4]{Physical Science and Engineering Division, King Abdullah University of Science and Technology (KAUST), Thuwal 23955-6900, Kingdom of Saudi Arabia}

\begin{document}

\maketitle

\begin{abstract}
  We propose a three-dimensional micromagnetic model that dynamically solves the Landau-Lifshitz-Gilbert equation coupled to the full spin-diffusion equation.
  In contrast to previous methods, we solve for the magnetization dynamics and the electric potential in a self-consistent fashion.
  This treatment allows for an accurate description of magnetization dependent resistance changes.
  Moreover, the presented algorithm describes both spin accumulation due to smooth magnetization transitions and due to material interfaces as in multilayer structures.
  The model and its finite-element implementation are validated by current driven motion of a magnetic vortex structure.
  In a second experiment, the resistivity of a magnetic multilayer structure in dependence of the tilting angle of the magnetization in the different layers is investigated.
  Both examples show good agreement with reference simulations and experiments respectively.
\end{abstract}
  
\newpage
\section{Introduction}
Spin-tronic devices are versatile candidates for a variety of applications including sensors \cite{daughton1999gmr,freitas2007magnetoresistive}, storage devices \cite{huai2008spin}, and frequency generators \cite{kiselev2003microwave,mistral2006current}.
Different quantum mechanical mechanisms contribute to the coupling of the electrical current to the magnetization.
Simulations of spintronic systems usually apply the micromagnetic model extended by simplified coupling terms such as the model by Slonczeswki \cite{slonczewski2002currents} and the model by Zhang and Li \cite{zhang2004roles}.
In \cite{abert2015three} it was shown that a simplified spin diffusion model incorporates these models.
In all these approaches, the effects of the magnetic state of the system on the electronic transport are neglected.
Indeed, the electric current density is assumed to be fixed, so that the models can be only used to investigate how the electronic transport affects the magnetization dynamics, but not vice versa.
In this work, we present a three-dimensional finite-element method for the solution of the full spin diffusion model from \cite{zhang2002mechanisms} which includes a bidirectional coupling of the magnetization to the electric current.
In \cite{sturma2015geometry} the model from \cite{zhang2002mechanisms} is applied to a single phase magnetic nanostructure in order to predict domain wall motion.
In this work we extend this model to composite structures consisting of magnetic and nonmagnetic materials which enables us to compute the magnetization-dependent resistivity caused by the GMR effect.

\section{The Model}\label{sec:model}
\begin{figure}[b]
  \centering
  \includegraphics{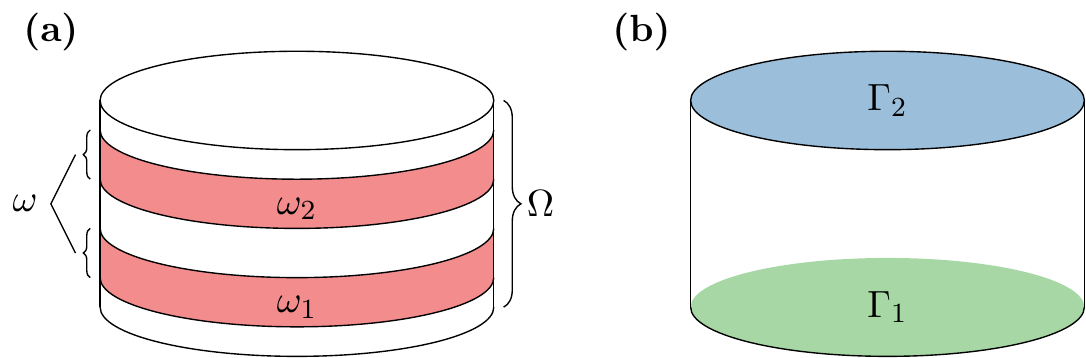}
  \caption{
    Typical magnetic/nonmagnetic material stack as used for MRAM devices.
    (a) The region $\omega = \omega_1 \cup \omega_2$ denotes the magnetic material, while $\Omega$ denotes the complete sample.
    (b) Electrical contacts $\Gamma_1$ and $\Gamma_2$.
  }
  \label{fig:regions}
\end{figure}
According to the micromagnetic model, the magnetization dynamics in a three-dimensional magnetic domain $\omega$ is described by the Landau-Lifshitz-Gilbert equation (LLG)
\begin{equation}
  \frac{\partial \vec{m}}{\partial t} =
  - \gamma \vec{m} \times \left( \vec{h}_\text{eff} + \frac{J}{\hbar \gamma M_\text{s}} \vec{s} \right)
  + \alpha \vec{m} \times \frac{\partial \vec{m}}{\partial t}
  \quad \text{in} \quad \omega,
  \label{eq:llg}
\end{equation}
where $\vec{m}$ is the normalized magnetization, $\gamma$ is the gyromagnetic ratio, $\alpha$ is the Gilbert damping, and $\vec{h}_\text{eff}$ is the effective field that usually contains the demagnetization field, the exchange field, as well as other contributions depending on the problem setting.
The effective field is complemented by a contribution from the spin accumulation $\vec{s}$ with $J$ being the exchange strength between itinerant and localized spins, $\hbar$ being the reduced Planck constant, and $M_\text{s}$ being the saturation magnetization.
The spin accumulation itself is defined in the conducting region $\Omega$ and satisfies the equation of motion
\begin{equation}
  \frac{\partial \vec{s}}{\partial t} =
  - \vnabla \cdot \vec{j}_\text{s}
  - \frac{\vec{s}}{\tau_\text{sf}} - J \frac{\vec{s} \times \vec{m}}{\hbar}
  \quad \text{in} \quad \Omega,
  \label{eq:spin_accumulation}
\end{equation}
where $\tau_\text{sf}$ is the spin-flip relaxation time, and $\vec{j}_\text{s}$ is the matrix-valued spin current.
According to \cite{zhang2002mechanisms, imamura2011spin}, the spin current $\vec{j}_\text{s}$ and the electric current $\vec{j}_\text{e}$ are defined by
\begin{align}
  \vec{j}_\text{e} &= 2 C_0 \vec{E} - 2 \beta' D_0 \frac{e}{\mu_\text{B}} (\vnabla\vec{s})^T \vec{m} \label{eq:electric_current}, \\
  \vec{j}_\text{s} &= 2 \beta C_0 \frac{\mu_\text{B}}{e} \vec{m} \otimes \vec{E} - 2 D_0 \vnabla\vec{s}, \label{eq:spin_current}
\end{align}
where $\vec{E}$ is the electric field, $D_0$ is a diffusion constant, $C_0$ is related to electric resistivity $\rho$ by $C_0 = 1 / 2\rho$, and $\beta$ and $\beta'$ are dimensionless polarization parameters.
Solving \eqref{eq:electric_current} for $\vec{E}$ and inserting this into \eqref{eq:spin_current} yields
\begin{equation}
  \vec{j}_\text{s} =
  \beta \frac{\mu_\text{B}}{e} \vec{m} \otimes \vec{j}_\text{e}
  - 2 D_0 \left[
    \vnabla\vec{s}
    - \beta \beta' \vec{m} \otimes ( (\vnabla\vec{s})^T \vec{m})
  \right],
  \label{eq:spin_current_simple}
\end{equation}
which, combined with \eqref{eq:spin_accumulation}, gives the simplified diffusion model with prescribed electric current $\vec{j}_\text{e}$ used in \cite{garcia2007spin,ruggeri2015coupling,abert2015three}.

However, instead of prescribing the electric current, it is possible to solve the coupled system \eqref{eq:electric_current} and \eqref{eq:spin_current}.
For this purpose, we consider the following simplifications:
As proposed in \cite{ruggeri2015coupling}, we assume the spin accumulation to be in equilibrium at all times, i.e.,
\begin{equation}
  \frac{\partial \vec{s}}{\partial t} = 0.
\end{equation}
This simplification is justified by the fact that the characteristic time scale of the spin accumulation dynamics is two orders of magnitude smaller than that of the magnetization dynamics, see \cite{zhang2004roles}.
Since sample sizes are usually very small, eddy currents can be neglected \cite{hrkac2005influence}.
Therefore, the electric field is curl free and thus given as the gradient of a scalar potential
\begin{equation}
  \vec{E} = - \vnabla u.
\end{equation}
Moreover, it is assumed that the conducting region $\Omega$, that is considered for the solution of the system \eqref{eq:electric_current} and \eqref{eq:spin_current}, does not contain any sources of electric currents, i.e.,
\begin{equation}
  \vnabla \cdot \vec{j}_\text{e} = 0.
\end{equation}
Inserting these assumptions into \eqref{eq:spin_accumulation} -- \eqref{eq:spin_current} yields the overall system
\begin{align}
  - 2 \, \vnabla \cdot \left[
      C_0 \vnabla u + \beta' D_0 \frac{e}{\mu_\text{B}} (\vnabla\vec{s})^T \vec{m}
  \right] &= 0, \label{eq:system_u}\\
  \vnabla \cdot \left[
     2 \beta C_0 \frac{\mu_\text{B}}{e} \vec{m} \otimes \vnabla u
     + 2 D_0 \vnabla\vec{s}
  \right]
  - \frac{\vec{s}}{\tau_\text{sf}} - J \frac{\vec{s} \times \vec{m}}{\hbar}
  &= 0, \label{eq:system_s}
\end{align}
that determines the electric potential $u$ and the spin accumulation $\vec{s}$.

A number of boundary conditions are required in order to close the LLG \eqref{eq:llg} coupled to the spin-diffusion system for the magnetization $\vec{m}(t)$.
The LLG itself is an initial value problem and requires the initial magnetization
\begin{equation}
  \vec{m}(t_0) = \vec{m}_0.
\end{equation}
If the exchange field $\vec{h}_\text{exchange} = 2A/(\mu_0 M_\text{s}) \Delta \vec{m}$, with the exchange constant $A$ and the saturation magnetization $M_\text{s}$, is included in the list of effective field contributions, an additional boundary condition for the magnetization is required.
If the domain $\omega$ for the solution of the LLG coincides with the magnetic region, it was shown in \cite{rado1959spin} that homogeneous Neumann boundary conditions are the right choice in a physical sense
\begin{equation}
  \frac{\partial \vec{m}}{\partial \vec{n}} = 0
  \quad \text{on} \quad \partial \omega.
\end{equation}
The system \eqref{eq:system_u} -- \eqref{eq:system_s} introduces the need for further boundary conditions for the electric potential $u$ and the spin accumulation $\vec{s}$.
A set of mixed boundary conditions is used to prescribe the electric potential and current inflow at the boundary of the conducting region $\partial \Omega$.
The Dirichlet condition is applied directly to the potential $u$
\begin{equation}
  u = u_0 \quad \text{on} \quad \Gamma_{D} \subseteq \partial \Omega,
\end{equation}
while the Neumann condition is applied to the electric current
\begin{equation}
  \vec{j}_\text{e} \cdot \vec{n} =
  - 2 \left[
  C_0 \vnabla u + \beta' D_0 \frac{e}{\mu_\text{B}} \left[ (\vnabla\vec{s})^T \vec{m} \right]
  \right] \cdot \vec{n}
  =
  g
  \quad \text{on} \quad \Gamma_\text{N} = \partial \Omega \setminus \Gamma_\text{D}.
\end{equation}
A typical choice of these boundary conditions is depicted in Fig.~\ref{fig:regions}(b).
In an MRAM like structure, the top and bottom surfaces $\Gamma_1$ and $\Gamma_2$ are electric contacts. 
In order to prescribe the current flow through the sample, like it is usually done in experiments, one might set the potential to zero at one contact $u = 0$ on $\Gamma_1$.
On the other contact $\Gamma_2$ a finite current inflow is prescribed $\vec{j}_\text{e} \cdot \vec{n} = g$.
The rest of the sample boundary $\partial \Omega \setminus (\Gamma_1 \cup \Gamma_2)$ is treated with homogeneous Neumann conditions $\vec{j}_\text{e} \cdot \vec{n} = 0$.

The boundary conditions are completed with homogeneous Neumann conditions for the spin accumulation
\begin{equation}
  \vnabla \vec{s} \cdot \vec{n} = 0
  \quad \text{on} \quad \partial \Omega.
  \label{eq:s_neumann}
\end{equation}
This choice can be justified by considering the boundary flux of the spin current.
Multiplying~\eqref{eq:spin_current_simple} with the boundary normal $\vec{n}$ and inserting the Neumann condition yields
\begin{align}
  \vec{j}_\text{s} \cdot \vec{n} &=
  \beta \frac{\mu_\text{B}}{e} \vec{m} (\vec{j}_\text{e} \cdot \vec{n})
  - 2 D_0 \left[
    \vnabla\vec{s} \cdot \vec{n}
    - \beta \beta' \vec{m} ((\vnabla\vec{s} \cdot \vec{n}) \cdot \vec{m})
  \right]\\
  &= \beta \frac{\mu_\text{B}}{e} \vec{m} (\vec{j}_\text{e} \cdot \vec{n})
\end{align}
This expression is nonzero only at boundaries with both nonvanishing current in-/outflow and nonvanishing magnetization.
Hence the homogeneous Neumann condition \eqref{eq:s_neumann} is equivalent to the noflux condition $\vec{j}_\text{s} \cdot \vec{n} = 0$ for systems as depicted in Fig.~\ref{fig:regions}, where the electric contacts are part of the nonmagnetic region.
The noflux condition itself is a reasonable choice if the thickness of the electrodes is large against the diffusion length.
In this case the spin accumulation and hence also the spin current is expected to be approximately zero at the contacts.

For systems where the magnetic region is contacted directly, the homogeneous Neumann condition leads to unphysical behaviour since the spin accumulation that is generated at the contact interface is neglected.
This accumulation strongly depends on the material of the leads that is not known when directly contacting the magnet.
However, the choice of homogeneous Neumann conditions leads to a good agreement with the predictions of the model by Zhang and Li \cite{zhang2004roles} that also neglects surface effects at the contacts.

\section{Validation}\label{sec:sp5}
The presented model is implemented within the finite-element code magnum.fe \cite{abert2013magnum}.
The discretization is explained in detail in Appendix~\ref{sec:discretization}.
For validation purposes, the standard problem \#5 proposed by the \textmu MAG group \cite{mumag5} is computed with the self-consistent model and compared to results obtained with the model of Zhang and Li \cite{zhang2004roles} and the simplified diffusion model used in \cite{abert2015three}.
\begin{figure}
  \includegraphics{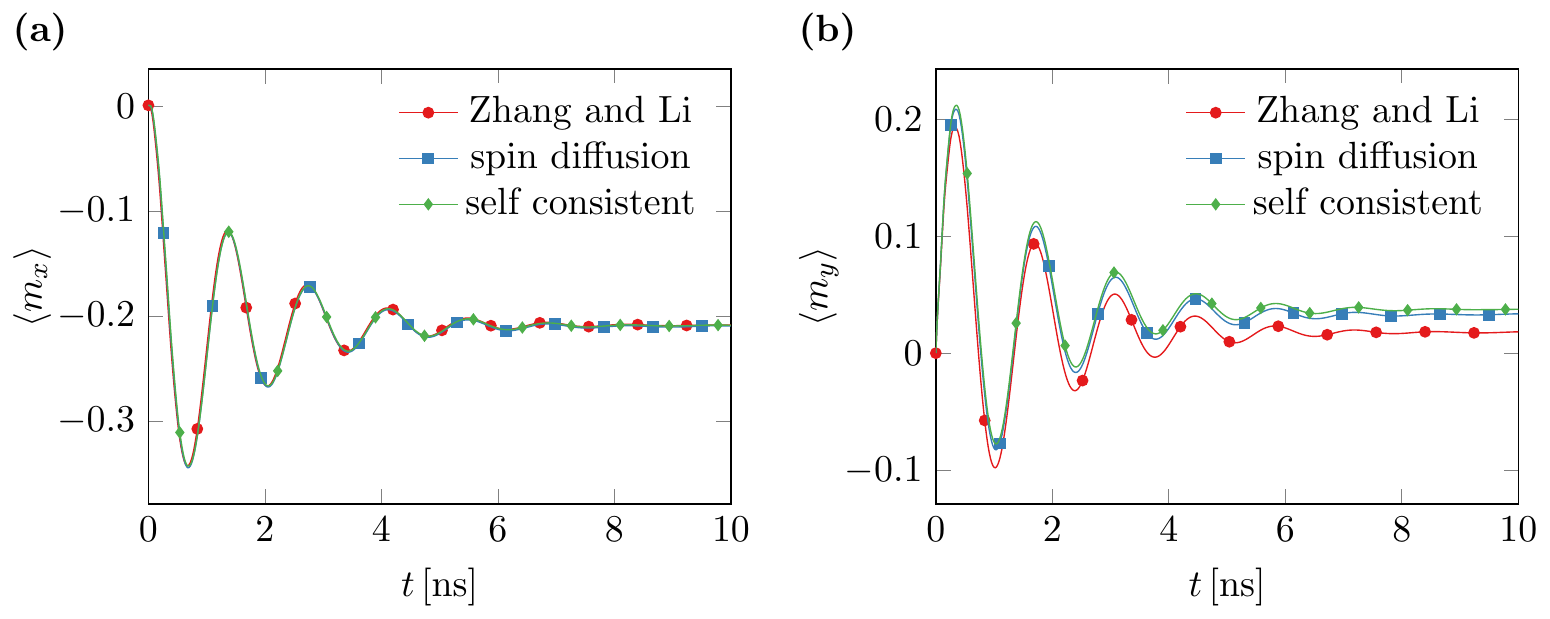}
  \caption{
    Results for the standard problem \#5 for different micromagnetic models.
    Time evolution of the averaged magnetization components.
    (a) $x$-component
    (b) $y$-component.
  }
  \label{fig:sp5}
\end{figure}
While this problem does not require particular features of the proposed self-consistent model, it serves as an excellent experiment for the validation of the proposed algorithm.
The standard problem \#5 describes the motion of a magnetic vortex in a thin square of size \SI{100 x 100 x 10}{\nano\metre} under the influence of a DC current defined by $\beta \vec{j}_\text{e} = (10^{12}, 0, 0)\,\si{\ampere/\metre^2}$.
For our simulations we choose $\beta = 1$ and thus $\vec{j}_\text{e} = (10^{12}, 0, 0)\,\si{\ampere/\metre^2}$.
The material parameters are chosen similar to those of permalloy, namely $M_s = \SI{8e5}{\ampere/\metre}$, $A = \SI{1.3e-11}{\joule/\metre}$, and $\alpha = 0.1$.
In the original problem definition, it is proposed to apply the model of Zhang and Li that extends the LLG \eqref{eq:llg} by current dependent terms
\begin{equation}
  \frac{\partial \vec{m}}{\partial t} =
  - \gamma \vec{m} \times \vec{h}_\text{eff}
  + \alpha \vec{m} \times \frac{\partial \vec{m}}{\partial t}
  - b \vec{m} \times [\vec{m} \times (\vec{j}_\text{e} \cdot \vnabla) \vec{m}]
  - \xi b \vec{m} \times (\vec{j}_\text{e} \cdot \vnabla) \vec{m},
\end{equation}
where $b = \SI{72.17e-12}{\metre^3 \per(\ampere\second)}$ is a coupling constant and $\xi = 0.05$ the degree of nonadiabacity.
As shown in \cite{abert2015three} an equivalent set of material parameters for the diffusion model can be obtained by perceiving the Zhang and Li model as diffusion model in the limit of vanishing diffusion.
For the diffusion model we choose $D_0 = \SI{e-3}{\square\metre/\second}$, $\beta' = 0.8$, and $\tau_\text{sf} = \SI{5e-14}{\second}$.
The remaining constant $J = \SI{0.263}{\electronvolt}$ is then uniquely defined by the relations given in \cite{abert2015three}.
In order to solve this problem with the self-consistent model the additional conductivity constant $C_0 = \SI{1.2e6}{\ampere\per(\volt\metre)}$ is introduced.
Furthermore, instead of prescribing a constant current within the magnetic material, the current is applied in terms of boundary conditions.
On the left side of the sample $\Gamma_\text{N} = \{ \vec{r} | r_x = \SI{-50}{\nano\meter}\}$ current inflow is set to $\vec{j}_\text{e} \cdot \vec{n} = \SI{e12}{\ampere\per\metre^2}$ and on the right side of the sample $\Gamma_\text{D} = \{ \vec{r} | r_x = \SI{+50}{\nano\meter}\}$ the potential is set to $0$.
The remaining boundary is treated with homogeneous Neumann conditions in order to simulate current in- and outflow only through the contacts.

The results for the computation of standard problem \#5 with the different methods are shown in Fig.~\ref{fig:sp5}.
While the results for the averaged $x$-component of the magnetization are in very good agreement, the results for the $y$-component show a notable offset.
The offset of the results of the Zhang and Li model to the remaining models is caused by the neglected diffusion.
The offset of the self-consistent model to the simple diffusion model is caused by the inhomogeneous current distribution resulting from the self-consistent treatment.

\section{Resistance of multilayer stack with perpendicular current}
The key advantage of the presented method over the simplified diffusion model introduced in \cite{abert2015three} is the self-consistent treatment of the electric potential $u$.
The potential is computed considering not only Ohm's law $\vec{j}_\text{e} = 2 C_0 \vec{E}$ but also magnetization dependent contributions.
This dependence is exploited for instance in sensor applications \cite{daughton1999gmr}.
Consider a magnetic multilayer stack as shown in Fig.~\ref{fig:regions} with two homogeneously magnetized layers $\omega_1$ and $\omega_2$ separated by a conducting layer.
The resistivity of this structure heavily depends on the tilting angle $\theta$ of the magnetization in $\omega_1$ and $\omega_2$.
Namely an antiparallel configuration is known to result in a high resistivity while a parallel configuration leads to a low resistivity.
This effect is referred to as giant magnetoresistance (GMR).
In order to calculate the electric resistivity with the presented model, the potential difference between the contacts $\Gamma_1$ and $\Gamma_2$ is computed for a given current inflow.

\begin{figure}
  \centering
  \includegraphics{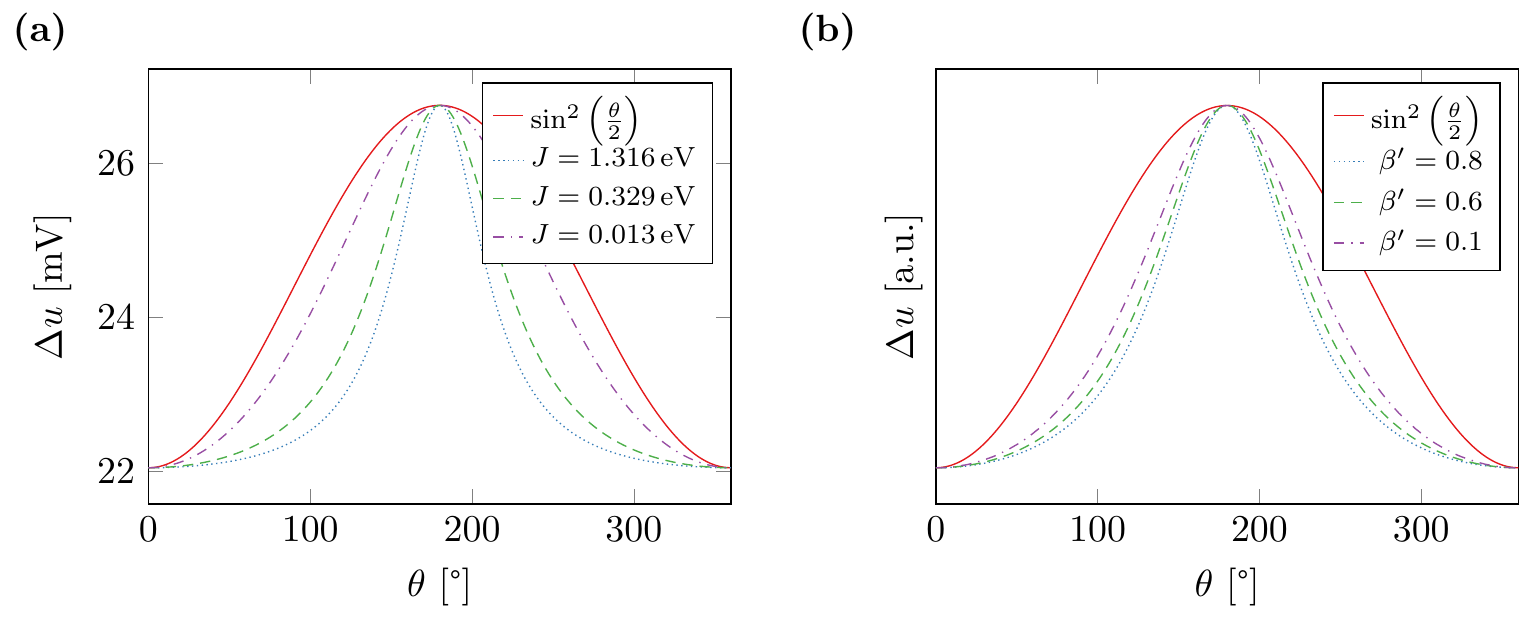}
  \caption{
    Potential difference $\Delta u$ between top and bottom contact required to generate an average current density of $j_\text{e} = \SI{e12}{\ampere/\metre^2}$ depending on the tilting angle of the magnetization in the free and fixed layer of a magnetic multilayer structure.
    (a)~variation of $J$ for $\beta' = 0.8$
    (b)~variation of $\beta'$ for $J = \SI{0.263}{\electronvolt}$.
    The results are renormalized in order to facilitate the comparison to the sine parameterization.
  }
  \label{fig:gmr}
\end{figure}
For numerical experiments two magnetic layers with \SI{5}{\nano\metre} thickness separated by a conducting layer with \SI{1.5}{\nano\metre} thickness are considered.
The system is contacted with \SI{100}{\nano\metre} thick leads in order to justify the homogeneous boundary conditions on the spin accumulation as described in Sec.~\ref{sec:model}.
The potential is set to $u = 0$ at the bottom contact $\Gamma_1$ and the current inflow is set to $\vec{j}_\text{e} \cdot \vec{n} = \SI{e12}{\ampere/\metre^2}$ on the top contact $\Gamma_2$.
Note that the cross section of the system does not have any influence on the potential computation as long as it is constant throughout the stack.

Fig.~\ref{fig:gmr} shows the computed potential difference for different tilting angles of the magnetization in the two layers $\omega_1$ and $\omega_2$.
The material parameters in the magnetic regions are chosen similiar to those in Sec.~\ref{sec:sp5}.
In the conducting region $\Omega \setminus \omega$, a different diffusion constant of $D_0 = \SI{5e-3}{\metre^2/\second}$ and a conductivity of $C_0 = \SI{6.0e6}{\ampere/(\volt\metre)}$ is applied.
Moreover, in Fig.~\ref{fig:gmr}(a), exchange strength $J$ is varied in the whole stack $\Omega$.
In Fig.~\ref{fig:gmr}(b), the polarization parameter $\beta'$ is varied.
The resulting potential is compared to a sine parameterization $a+b\sin^2(\theta/2)$ that is often used to describe the GMR effect in such a stack \cite{dieny1991giant} in the presence of some in-plane current.
\begin{figure}
  \centering
  \includegraphics{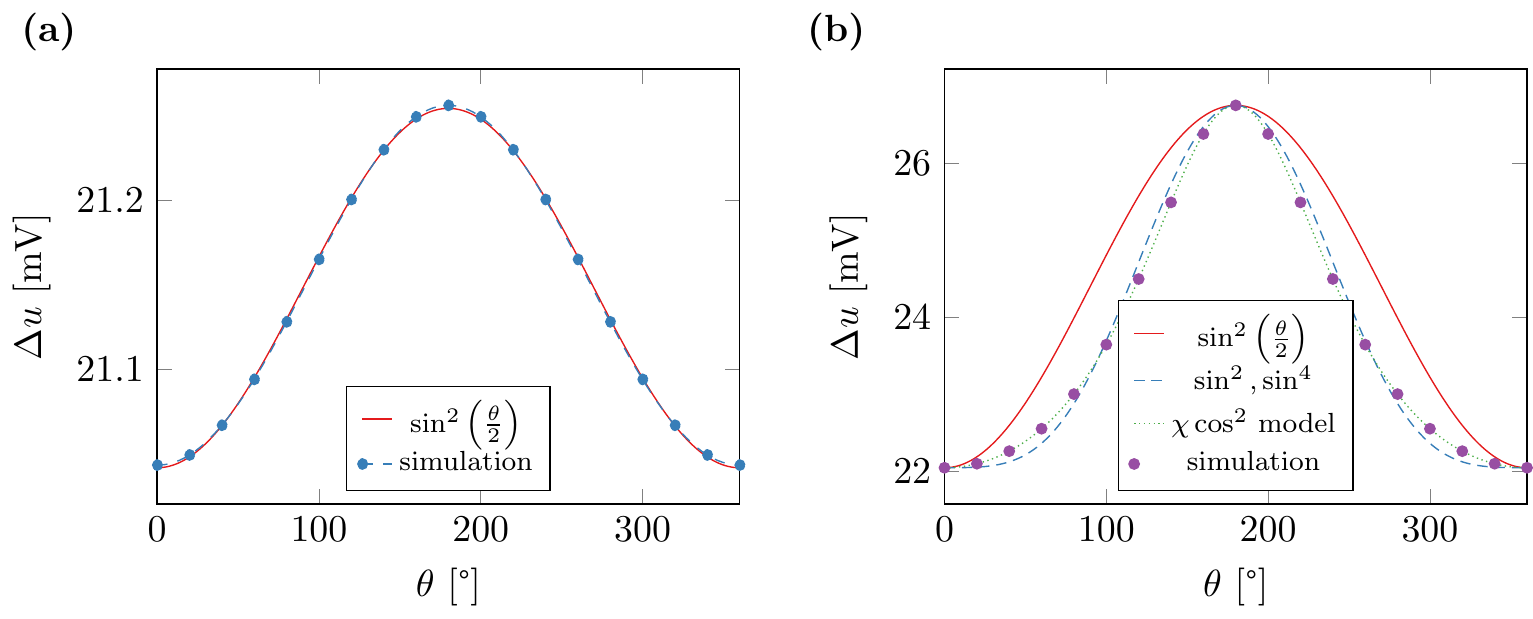}
  \caption{
    Potential difference $\Delta u$ between top and bottom contact required to generate an average current density of $j_\text{e} = \SI{e12}{\ampere/\metre^2}$ depending on the tilting angle of the magnetization in the free and fixed layer of a magnetic multilayer structure along with fits to different models for
    (a) $J = \SI{0.013}{\electronvolt}$ and $\beta' = 0.1$
    (b) $J = \SI{0.082}{\electronvolt}$ and $\beta' = 0.8$.
  }
  \label{fig:gmr_sine}
\end{figure}
The presented simulations however suggest that the potential and thus the resistivity of the stack out-of-plane currents is not well described by a sine, but has a much narrower peak for certain choices of material parameters.
Specifically the sine approximation is accurate only in the case of small $\beta'$ and $J$ as shown in Fig.~\ref{fig:gmr_sine}(a), where the parameters are chosen as $\beta' = 0.1$ and $J = \SI{0.013}{\electronvolt}$.

The deviation of the angular dependence of the resistivity from the sine approximation for some out-of-plane current was already observed in experiment \cite{dauguet1996angular}.
The authors of this work suggest to include a higher order sine term in order to fit the resistivity curve.
Fig.~\ref{fig:gmr_sine}(b) shows the result of the fit with $a + b\sin^2(\theta/2) + c\sin^4(\theta/2)$ for $J = \SI{0.082}{\electronvolt}$ and $\beta' = 0.8$ which shows a good agreement with the simulated curve.
A similar effect is also predicted in \cite{slonczewski2002currents}, where the authors suggest the following expression for the angular dependence of the resistivity
\begin{equation}
  R = \frac{1 - \cos^2(\theta/2)}{1 + \chi \cos^2(\theta/2)}.
  \label{eqn:r_slonczewski}
\end{equation}
The parameter $\chi$ depends on geometry and material of the involved layers and leads to a steeper peek if positive, which, according to the reference, is the case for all investigated systems up to then.
Using $\chi$ as a fitting parameter, the results for $J = \SI{0.082}{\electronvolt}$ and $\beta' = 0.8$ can be described with very high precision, see Fig.~\ref{fig:gmr_sine}(b).

The same effect was also predicted theoretically in \cite{stiles2002noncollinear} with a model of Valet and Fert \cite{valet1993theory} and in \cite{strelkov2011spin} with a two-dimensional diffusion model.
However, these papers do not discuss the influence of material parameters onto the narrowing of the sine response in detail.

In the context of the diffusion model, the deviation from the simple sine approximation $\sin^2(\theta/2)$ has its origin in two different effects.
First, the cross product term $J \vec{s} \times \vec{m} / \hbar$ in \eqref{eq:system_s} describes the torque that is exerted from the magnetization on the spin polarization of the itinerant electrons.
This torque is zero for parallel and antiparallel alignment of the magnetic layers and reduces the angle of the polarization of the itinerant electrons to the magnetization for any other alignment.
Hence, for large $J$ this contribution leads to a lowered resistance for all alignments other than parallel and antiparallel, which results in the narrow peak observed in the simulation.

The second effect is a bit more subtle.
For vanishing $\beta'$ the potential $u$ from \eqref{eq:system_u} varies linearly.
Small values of $\beta'$ lead to small perturbations of this linear solution.
While these perturbations have a clear effect on the overall potential difference, their effect on the spin accumulation $\vec{s}$ due to \eqref{eq:system_s} is negligible, leading to a clean sinusoidal response of the system as shown in Fig.~\ref{fig:gmr_sine}(a).
With increasing $\beta'$ the perturbations of $u$ gain influence on the solution of $\vec{s}$ which results in a distorsion of the sinusoidal response as seen in Fig.~\ref{fig:gmr} and \ref{fig:gmr_sine}.

\section{Conclusion}
We propose a three-dimensional spin-diffusion model that simultaneously solves for the spin accumulation $\vec{s}$ and the electric potential $u$.
By coupling this model to the Landau-Lifshitz-Gilbert equation, we are able to self-consistently solve the magnetization dynamics for a given current inflow.
In order to validate the model and its implementation, we simulate the standard problem \#5 proposed by the \textmu MAG group and compare the outcome to results obtained with simplified models.

In a second numerical experiment, we compute the resistivity of a magnetic multilayer structure in dependence on the tilting angle of the magnetization in the two magnetic layers.
In the limit of small polarization parameter $\beta'$ and a small exchange strength $J$, we show that the resistivity is well approximated by a sine.
For realistic choices of $\beta'$ and $J$ the angular dependence shows a significantly narrower peak than the simple sine approximation.
While existing models already predict the observed behaviour in a macro spin approach, the presented model is able to accurately describe GMR effects for both dynamically and spatially varying magnetization configurations.

\appendix
\section{Discretization}\label{sec:discretization}
We solve the coupled equations \eqref{eq:llg} and \eqref{eq:system_u} -- \eqref{eq:system_s} numerically by employing the finite-element method for spatial discretization and a preconditioned BDF scheme as described in \cite{suess2002time} for the time integration.
The demagnetization field is solved by a hybrid FEM--BEM method as introduced in \cite{fredkin1990hybrid}.
For the discretization of \eqref{eq:system_u} and \eqref{eq:system_s} special care has to be taken in the treatment of the discontinuities, e.g., in the magnetization $\vec{m}$ introduced by magnetic--nonmagnetic interfaces.
As usual the original problem is multiplied with test functions and integration by parts is applied to avoid second derivatives.
Furthermore, first derivatives of the magnetization $\vec{m}$ are eliminated in the same way and the integration domain for terms including the magnetization is restricted to the magnetized region $\omega$ in order account for discontinuities.

The solution variables $\vec{m}$, $u$ and $\vec{s}$ as well as the test functions $v$ and $\vec{\zeta}$ are discretized by (componentwise) piecewise affine, globally continuous functions constructed on a tetrahedral mesh.
The material parameters $\beta$, $\beta'$, $C_0$, $D_0$, $\tau_\text{sf}$, and $J$ are discretized with piecewise constant functions.
For a given magnetization $\vec{m}$, the weak version of \eqref{eq:system_u} reads
\begin{equation}
  \int_\Omega 2 C_0 \vnabla u \cdot \vnabla v \,\text{d}\vec{x}
  + \int_\omega 2 \beta' D_0 \frac{e}{\mu_\text{B}} (\vnabla\vec{s})^T \vec{m}
  \cdot \vnabla v \,\text{d}\vec{x}
  =
  - \int_{\Gamma_\text{N}} g v \,\text{d}\vec{s},
  \label{eq:discrete_u}
\end{equation}
where the current in-/outflow is given by $g$ as Neumann condition. 
The additional Dirichlet boundary conditions on $\Gamma_\text{D}$ are embedded in the function space of the solution $u$ as usual when employing the finite-element method.
The weak version of \eqref{eq:system_s} reads
\begin{multline}
  - \int_\omega 2 \beta C_0 \frac{\mu_\text{B}}{e} \vec{m} \otimes \vnabla u : \vnabla\vec{\zeta} \,\text{d}\vec{x}
  + \int_{\partial \omega \cap \Gamma_\text{D}} 2 \beta C_0 \frac{\mu_\text{B}}{e} (\vnabla u \cdot \vec{n})(\vec{m} \cdot \vec{\zeta}) \,\text{d}\vec{s}
  \\
  - \int_\Omega 2 D_0 \vnabla\vec{s} : \vnabla\vec{\zeta} \,\text{d}\vec{x}
  - \int_\Omega \frac{\vec{s} \cdot \vec{\zeta}}{\tau_\text{sf}} \,\text{d}\vec{x}
  - \int_\omega J \frac{(\vec{s} \times \vec{m}) \cdot \vec{\zeta}}{\hbar} \,\text{d}\vec{x} \\
  = 
  \int_{\partial \omega \cap \Gamma_\text{N}} 
  \beta \frac{\mu_\text{B}}{e} g (\vec{m} \cdot \vec{\zeta}) \,\text{d}\vec{s}.
  \label{eq:discrete_s}
\end{multline}

\section*{Acknowledgements}
The financial support by
the Austrian Federal Ministry of Science, Research and Economy and the National Foundation for Research, Technology and Development
as well as
the Austrian Science Fund (FWF) under grant W1245 and F4112 SFB ViCoM,
the Vienna Science and Technology Fund (WWTF) under grant MA14-44,
the innovative projects initiative of TU Wien
is gratefully acknowledged.
A.M. acknowledges financial support from the King Abdullah University of Science and Technology (KAUST).

\end{document}